\documentclass[a4paper,fleqn,usenatbib]{mnras}
\usepackage{mathptmx}
\usepackage{mathptmx}
\usepackage{txfonts}
\usepackage{txfonts}
\usepackage[T1]{fontenc}
\usepackage{ae,aecompl}
\usepackage{graphicx}	% Including figure files
\usepackage{amssymb}	% Extra maths symbols
\usepackage[dutch, english]{babel}
\usepackage{mathrsfs}
\usepackage{latexsym}
\usepackage{longtable}
\usepackage{multicol}
\usepackage{supertabular}
\usepackage{multirow}
\usepackage{epsf}
\usepackage{color}
\usepackage{float}
\usepackage{enumerate}
\usepackage{natbib}
\usepackage{arydshln}
\usepackage{comment}
\usepackage{capt-of}

\newcommand{\ms}{m~s$^{-1}$}
\newcommand{\um}{$\mu$m}
\newcommand{\deq}{d_{\rm eq}}
\newcommand{\dpre}{d_{\rm eq, pre}}
\newcommand{\dpost}{d_{\rm eq, post}}
\newcommand{\kgm}{kg~m$^{-3}$}

\newcommand{\mockalph}[1]{}

\title[]{The footprint of cometary dust analogues: \\ II. Morphology as a tracer of tensile strength and application to dust collection by the Rosetta spacecraft}
\author[L. E. Ellerbroek et al.]{
L. E. Ellerbroek$^{1}$\thanks{E-mail: ellerbroek@uva.nl},
B. Gundlach$^{2}$,
A. Landeck$^{2}$,
C. Dominik$^{1}$, 
J. Blum$^{2}$, 
\newauthor
S. Merouane$^{3}$,
M. Hilchenbach$^{3}$,
H. John$^{3}$
H. A. van Veen$^{4}$
\\
$^{1}$ Anton Pannekoek Institute for Astronomy, University of Amsterdam, Science Park 904, 1098 XH Amsterdam, The Netherlands \\
$^{2}$ Institut f\"{u}r Geophysik und extraterrestrische Physik, Technische Universit\"{a}t Braunschweig,\\ Mendelssohnstra\ss e 3, D-38106 Braunschweig, Germany\\
$^{3}$ Max-Planck-Institut f\"{u}r Sonnensystemforschung, Justus-von-Liebig-Weg 3, D-37077 G\"{o}ttingen, Germany\\
$^{4}$ Electron Microscopy Center Amsterdam, Academic Medical Center, Amsterdam, The Netherlands 
}

\date{Accepted XXX. Received YYY; in original form ZZZ}

\pubyear{2016}

\begin{document}
\label{firstpage}
\pagerange{\pageref{firstpage}--\pageref{lastpage}}
\maketitle

% Abstract of the paper
\begin{abstract}
The structure of cometary dust is a tracer of growth processes in the formation of planetesimals. Instrumentation on board the Rosetta mission to comet 67P/Churyumov-Gerasimenko captured dust particles and analysed them in situ. However, these deposits are a product of a collision within the instrument. 
We conducted laboratory experiments with cometary dust analogues, simulating the collection process by Rosetta instruments (specifically COSIMA, MIDAS). In Paper I we reported that velocity is a key driver in determining the appearance of deposits. 
Here in Paper II we use materials with different monomer sizes, and study the effect of tensile strength on the appearance of deposits. We find that mass transfer efficiency increases from $\sim1$ up to $\sim10$\% with increasing monomer diameter from 0.3~\um~to 1.5~\um~(i.e. tensile strength decreasing from $\sim 12$ to $\sim 3$~kPa), and velocities increasing from 0.5~to~6~\ms. Also, the relative abundance of small fragments after impact is higher for material with higher tensile strength. The degeneracy between the effects of velocity and material strength may be lifted by performing a closer study of the deposits. 
This experimental method makes it possible to estimate the mass transfer efficiency in the COSIMA instrument. Extrapolating these results implies that more than half of the dust collected during the Rosetta mission has not been imaged. 
We analysed two COSIMA targets containing deposits from single collisions. The collision that occurred closest to perihelion passage led to more small fragments on the target.
\end{abstract}

% Select between one and six entries from the list of approved keywords.
% Don't make up new ones.

%%%%%%%%%%%%%%%%%%%%%%%%%%%%%%%%%%%%%%%%%%%%%%%%%%

%%%%%%%%%%%%%%%%% BODY OF PAPER %%%%%%%%%%%%%%%%%%
%\clearpage
\section{Introduction}

Dust growth is the starting point of planet formation in the early solar system. However, various barriers exist that inhibit growth from dust to planetesimal, and onwards to larger bodies \citep{Dominik2007, Johansen2014, Blum2018}. On Earth, little or no geological remnants can be found from the earliest growth phases, as they have long since disappeared, having been heated and processed. Consequently, the most pristine remnants of the protosolar nebula surviving to this day are found in comets. Being kilometer-sized bodies that have spent most of their existence in cold regions of the solar system, they contain 'fossilized' evidence of early dust growth processes \citep{Blum2017}. 

This notion is supported by in-situ measurements of cometary dust particles by spacecraft \citep{Levasseur2018}. Most recently, the Rosetta mission to comet 67P/Churyumov-Gerasimenko (hereafter 67P) has provided a wealth of information on the  dust population in the coma. Two instruments on board Rosetta were able to image particles smaller than 1~mm: COSIMA ($10$~to a few 100~\um, \citealt{Kissel2007}) and MIDAS ($1$ to a few 10~\um, \citealt{Riedler2007}). These images show that dust particles in the coma consist of aggregates of dust grains with a hierarchical structure down to sub-micron scales \citep{Bentley2016, Mannel2016, Schulz2015}. Furthermore, various morphologies of dust aggregates were found (G\"{u}ttler et al., submitted to A\&A). Discussion is ongoing whether these morphologies relate to different dust species \citep{Langevin2016, DellaCorte2015, Fulle2015, Fulle2016b, Merouane2016, FulleBlum2017}. However, all of these experiments only look at the particles after they interacted with the spacecraft during collection, and therefore we do not necessarily have sufficient information on the particles before they were collected by the spacecraft.

One key line of research to interpret cometary dust data from space missions is therefore to perform laboratory experiments with cometary analogue materials. These allow to better understand dust growth processes, and moreover can interpret data from specific instruments where interaction with the equipment naturally influences the outcome of an in-situ measurement. Specifically, in the case of the COSIMA and MIDAS instruments, the dust particles imaged are the product of a collision with the instrument target surface, and possibly before that with the collection funnel. Experiments that simulate these collisions provide a tool for interpreting the spacecraft data.

Silicate aggregates provide a suitable analogue for cometary material after the ices have been sublimated. Their collision physics have been studied extensively both theoretically and experimentally. Experiments have been conducted both with polydisperse and monodisperse material \citep[for an overview, see][]{Blum2018, BlumWurm2008, Guttler2010}. However, up to this point studies have mostly focused on aggregate growth, and less on the part of the particle left on a solid surface after a collision. 

We performed laboratory experiments that aim to relate the properties of cometary dust analogues to the deposits they leave on a solid surface after impact. The aim of the first series of experiments, presented in \citet[][hereafter `Paper I']{Ellerbroek2017}, was to relate deposit morphology to particle size and velocity. In Paper I, a single polydispersed material was used as an analogue for cometary dust. A key insight gained from that study is that during impacts, aggregates fragmented and a large fraction of mass was lost in the instrument. Also, the velocity was seen as the main driver of the appearance of deposits. At impact velocities below the breaking or fragmentation barrier ($\sim2$~\ms~for the material used), particles either stick to or bounce off the target, leaving respectively a single (undamaged) deposit or a shallow footprint of loose monomers on the surface. Above the fragmentation barrier, particles fragment and leave pyramid-shaped deposits. An important open question that remains is: how do material properties (density, packing, monomer size, composition) and the resulting tensile strength influence deposit morphologies? 

In this Paper II, we present a second series of experiments and study the combined effect of tensile strength and velocity on the appearance of deposits. We use three different types of silicate aggregates, consisting of monodisperse monomers of a single size, which directly relates to the aggregate's tensile strength \citep{Gundlach2018}. This allows us for the first time to quantify the amount of mass transferred to the target plate, as a function of velocity and tensile strength, in a parameter range overlapping with the Rosetta experiments.

The paper is structured as follows. In Sect.~\ref{sec:methods}, we describe the test material, experimental setup, and the analysis method used. In Sect.~\ref{sec:results}, we present the quantitative results relating to mass transfer and deposit characteristics as a function of (pre-collection) dust properties. In Sect.~\ref{sec:discussion}, we discuss our results in the context of cometary dust measurements made by the Rosetta spacecraft. 
\label{sec:cosima}

%(Can a compact particle produce a shallow footprint?) 

%\clearpage
%SECTION: METHODS
\section{Methods}
\label{sec:methods}

In this section, we describe the test material used, the experimental setup and subsequent data analysis. 

%FIGURE: MATERIALS
	\begin{figure}
	\includegraphics[width=\columnwidth]{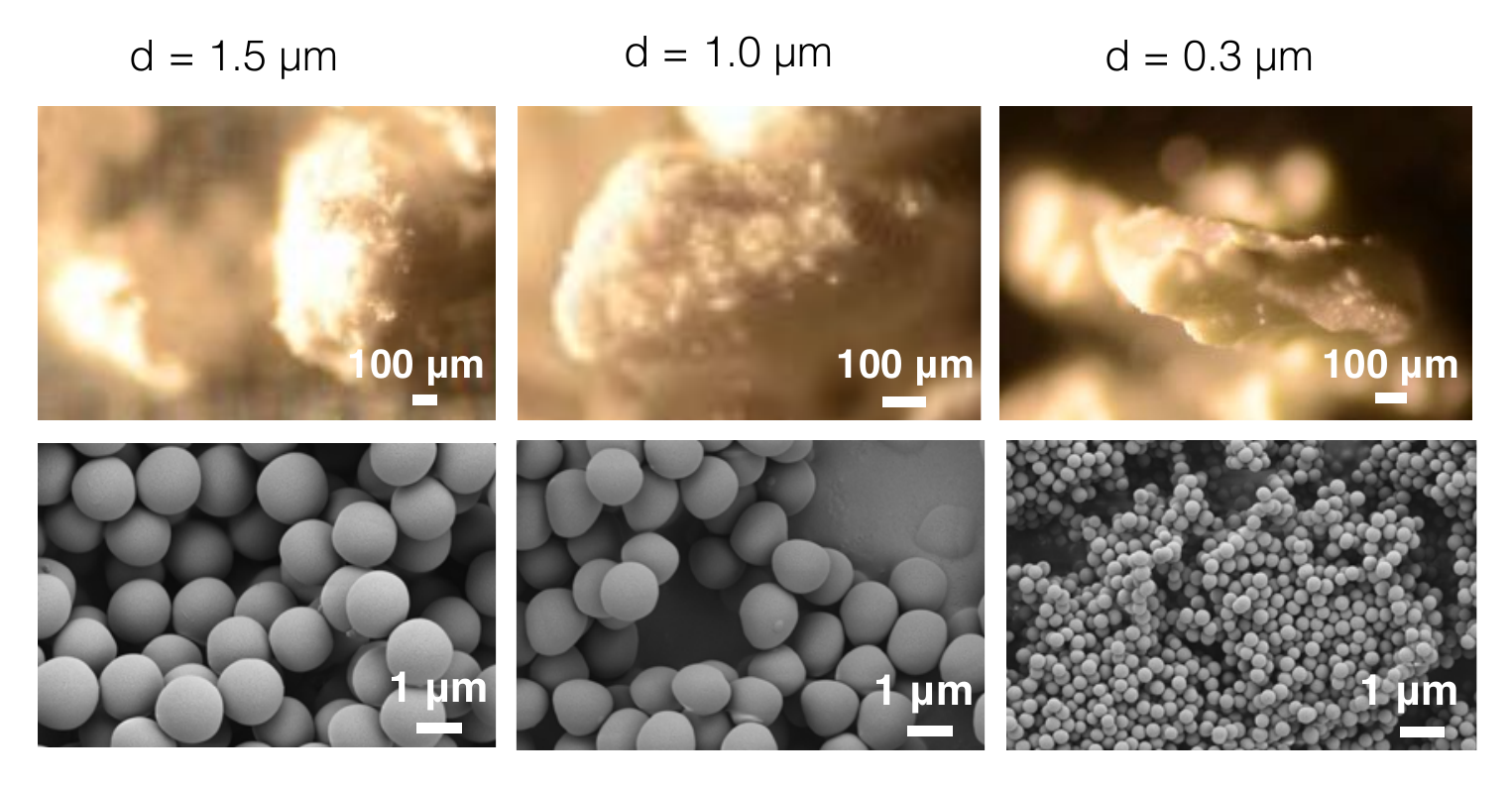}
    \caption{Optical microscope images (\textit{top}) and SEM images (\textit{bottom}) of the three different SiO$_2$ samples used in the experiments. The SEM images show the individual monomers while the optical images show the outside of a large aggregate (particle) as they form naturally in the storage container.}
    	\label{fig:samples}
	\end{figure}
	
	% TABLE: MATERIALS
\begin{table}
	\centering
	\caption{Characteristics of the three analogue particles used in this study. Here, $d_0$, $\phi$ and $\sigma$ are the diameter of the monomer grains, the packing density of the aggregates and the tensile strength of the aggregates, respectively. The latter are derived from \citet{Gundlach2018}.}
	\label{tab:materials}
	\begin{tabular}{cccc} % four columns, alignment for each
		\hline
		$d_0$ (\um) & $\phi$  & $\rho_{\rm b}$~($10^{3}$~\kgm) & $\sigma $ (kPa)\\
\hline
		1.5 & $0.35 \pm 0.03$  & $0.70 \pm 0.06$  & $2.5 \pm 0.4$\\ %1.8
		1.0 & $0.42 \pm 0.03$  & $0.84 \pm 0.06$ & $4.6 \pm 1.0$\\ %3.0
		0.3 & $0.32 \pm 0.02$  & $0.64 \pm 0.04$ & $11.6 \pm 1.5$ \\ %8.4
\hline
	\end{tabular}
\end{table}
	
\subsection{Test material}
\label{sec:methods:material}

The particles used in the experiments are aggregates of pure SiO$_2$, with a material density of $\rho_{\rm m}=2.0\times10^3$~\kgm. Three different samples\footnote{The materials were obtained from Micromod Partikeltechnologie GmbH under brand name Sicastar\textregistered; for technical specifications, see www.micromod.de} were used: aggregates of pure SiO$_2$ spherical monomers, with diameter $d_0=(0.3, 1.0, 1.5)$~\um, with a standard deviation of 4\%~\citep{Blum2006}. Optical microscope and scanning electron microscope (SEM) images are shown in Fig.~\ref{fig:samples}; the material properties are summarised in Tab.~\ref{tab:materials}. As the volume filling factor is of order $\phi \sim 0.3-0.4$, the resulting bulk density of these aggregates is $\rho_{\rm b} \equiv \rho_{\rm m} \phi = (0.7 \pm 0.1) \times 10^{3}$~\kgm. Note that the bulk density is not a function of monomer size, but rather scales with $\phi$. The material naturally forms aggregates in their storage canisters; we sieved these aggregates to obtain a particle diameter range of 100~--~400~\um, which overlaps the range in deposit sizes observed by COSIMA~\citep{Hornung2016}. 

An important parameter for the collision physics of the aggregates is the tensile strength~$\sigma$, which decreases as a function of monomer size. \citet{Gundlach2018} performed experiments to measure the tensile strength of the three types of material also used in this paper. Using the scaling relation they provide for different filling factors, we estimate the tensile strength of the materials to vary between  $\sim 3$~kPa and $\sim 12$~kPa for $d_0=1.5$~\um~to 0.3~\um~(Tab.~\ref{tab:materials}). This is also reflected by the shape of the aggregates that formed in the storage canisters: 1.5~\um~aggregates look more spherical (likely due to erosion while sieving) while aggregates of the 0.3~\um~material retain a more irregular shape even after sieving. However, the monomers themselves are smooth: \citet{Poppe2000a} measured surface roughness parameters of 0.5 and 1.2~\um~spheres, and find that their surfaces are smooth on a sub-nanometer scale. 

%FIGURE: SETUPCARTOON
	\begin{figure}
	\includegraphics[width=\columnwidth]{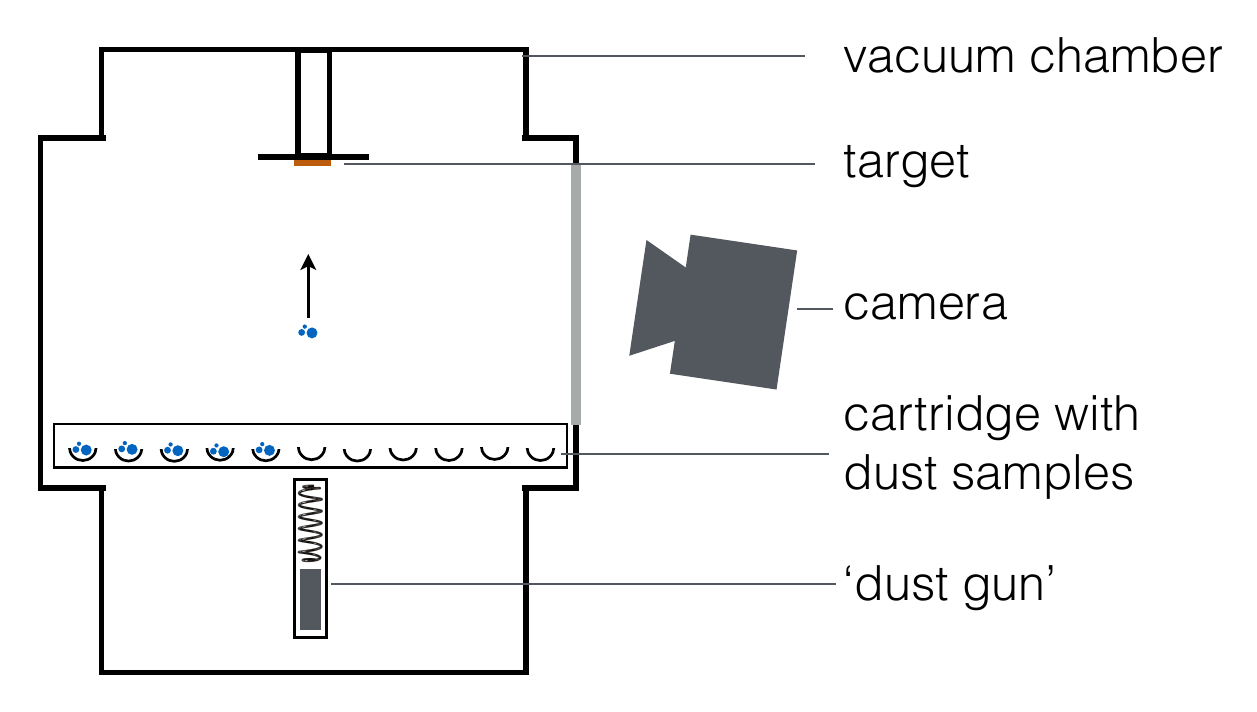}
    \caption{Schematic drawing of the experimental setup (side view). Taken from Paper I.}
    	\label{fig:setup cartoon}
	\end{figure}

\subsection{Experimental setup}
\label{sec:methods:experiment}

We use the same experimental setup as described in Paper~I. Here, we give a brief summary of the setup and some improvements that were made compared to Paper~I. 

A schematic drawing of the experimental setup is displayed in Fig.~\ref{fig:setup cartoon}. For each experiment, a collection of ~20 aggregates with diameter $100 - 400$~\um~were loaded onto a piston in a vacuum chamber. The chamber was depressurised to $\sim 0.03$~mbar, so as to minimise the influence of air drag. Although this pressure is much higher than the outer space environment where Rosetta operates, it suffices the purpose of minimising the influence of air drag on the collision time scales in the experiment. Similarly, the collision dynamics are not affected significantly by gravity (see Paper I, Fig.~4c). Our experiments thus approximate the circumstances in space, where contact forces are dominant in determining the collision dynamics.

A current pulse was applied to a lifting magnet, thereby launching the aggregates vertically onto a COSIMA target (polished silver) placed 8~cm above the piston. By tuning the voltage, impact velocities in the range 0.5~--~6.5~\ms~were reached, with a spread in velocity of $\sim$~0.5~\ms~within a single shot. Depending on its properties, a particle may stick, bounce or break up on the target surface, resulting in a collection of fragments left on the target surface. A collection of fragments resulting from the collision of a single particle is referred to as a `deposit'. 

The particle collisions were recorded by a high-speed camera, placed outside of the vacuum chamber. The camera was inclined at an angle of $6.7^\circ$, and was focused on the center of the target (focal depth 0.5~cm). The spatial resolution of the images was $20.7\pm0.3$~\um~per pixel. Simultaneously with launching, exposures of 0.05~ms were taken at a rate of 20,000 frames per second. In this way we fully monitored the particles' approach to and collisions on the target, allowing to measure their size, shape and velocity, and to monitor the collisions with the target surface. 

As described in Paper I, the average pressure exerted on an aggregate through acceleration during launch scales as $\langle P \rangle \propto v^2$, and is approximately 1~kPa for $v\sim1$~\ms. Considering the tensile strength of the different samples, particles can be expected to fragment upon launch when accelerated to a velocity above a fragmentation barrier of $(7.7,2.4,1.5)$~\ms~for monomer sizes $d_0=(0.3,1.0,1.5)$~\um. This expected breakup on launch as a function of monomer strength was indeed observed empirically (see Sect.~\ref{sec:results}), corroborating our earlier estimates of tensile strength. 

After launch and on-target collisions, the vacuum chamber was slowly pressurised. The target, containing deposits sticking to it, was dismounted with plastic tweezers only touching its sides, and photographed with a photo camera before it was stored for further analysis. This picture, taken immediately after the experiment, was used in later stages of the post-impact analysis as a reference to exclude any fragments added to or removed from the target plate during transfer, storage and further imaging (see Sect.~\ref{sec:methods:analysis}).

\subsection{Data analysis}
\label{sec:methods:analysis}

In this subsection, we describe the methods used to analyse the experimental data. The analysis was split into two parts: analysis of camera images (referred to as \textit{pre-impact}) and analysis of the resulting deposits on the targets (\textit{post-impact}). 

Our method differs somewhat from  Paper I, where we matched individual pre-impact and post-impact particles one-on-one. That method only allows studying a limited number of individual collisions. In the current study, we compare statistics of all particles in a single experiment pre-impact and deposits post-impact. This allows a quantitative measure of the mass transfer function, and bulk properties of the deposits. This situation also likely resembles the dust collection on the COSIMA plates, where a collection of fragments on target are often seen to be caused by a single pre-impact `parent' particle, which may already have been fragmented on the funnel wall shortly before hitting the target \citep{Merouane2017}.

We comment on the different post-impact deposit morphologies in Sect.~\ref{sec:results}.

\subsubsection{Pre-impact analysis }
\label{sec:methods:analysis:pre}

%FIGURE: PREIMPACT
	\begin{figure}
	\begin{center}
	\includegraphics[width=0.7\columnwidth]{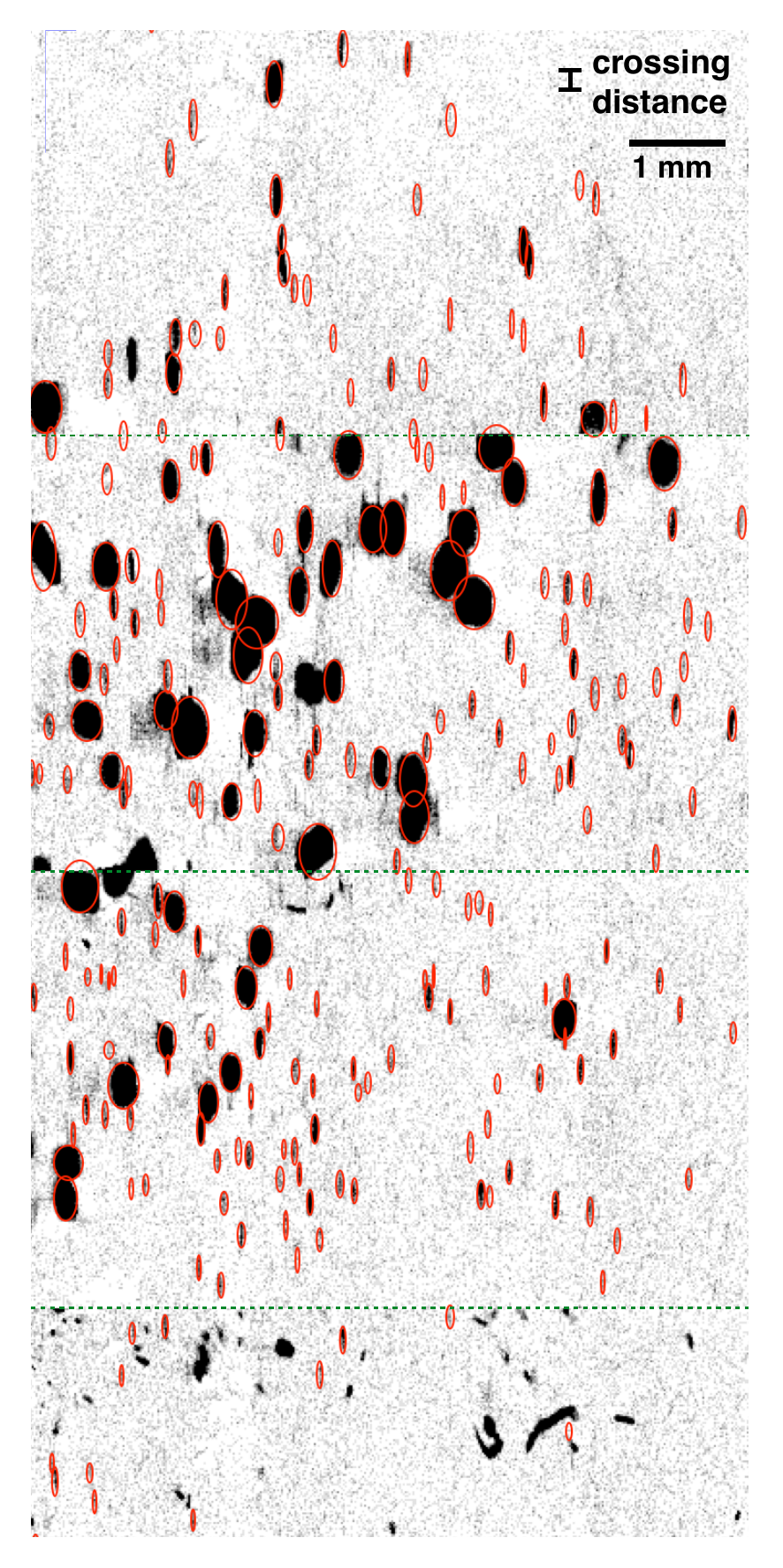}
	\end{center}
    \caption{Pre-impact mass measurement: montage of a cropped $0.5\times0.8$~cm area of 4 frames from a single experiment (A3, $v=5.4-5.6$~\ms). Time difference between frames is $\sim~0.5$~ms; exposure time of each frame is 0.05~ms. The frames are separated by green dashed lines, the time after launch increasing top to bottom. The target is located upwards of the frame. Particles with red ellipses drawn around them are moving upwards and eventually hit the target (ellipses that cross frames only pertain to one particle on one frame). The bottom frame is cropped as no more target-hitting particles were visible in the lower part of the frame. The bar in the top right corner indicates the crossing distance within a single exposure time. See Sect.~\ref{sec:methods:analysis:pre} for a detailed explanation of the method.}
    	\label{fig:preimpact}
	\end{figure}

Measuring pre-impact particle size distribution and mass was done by a careful examination of individual movie stills centered on the same location, at different times. A montage was made of 5~--~10 stills preceding the impact on target (see Fig.~\ref{fig:preimpact} for an example). The time difference between selected stills ranges from $\sim0.5$~ms (for high-velocity experiments) to $\sim4$~ms (for low-velocity experiments). The montage was constructed so that it displays all particles above the detection limit ($\dpre\gtrsim24$~\um) that eventually hit the target. Red ellipses are drawn around these particles (Fig.~\ref{fig:preimpact}). Unmarked particles either miss the target eventually, or are downward-moving rebounds off the target. 

As the particles cross multiple pixels during a single exposure, their silhouettes are `smeared out' in the vertical direction on the movie image. The size of each ellipse encompassing a particle was corrected for this blurring effect by reducing the vertical ellipse axis by the particle crossing distance (ranging from 2~--~14 pixels in the range $v=0.5-6.0$~\ms). Subsequently, the particle volume was approximated as 
\begin{equation}
V_{\rm pre}=\frac{\pi}{6}{\dpre}^3, 
\end{equation}
where $\dpre$~is the average diameter of the deconvolved ellipse. By inspection of multiple frames, we measured velocities and excluded particles that were moving downwards after rebounding.

This method results in the measurements, for every experiment, of the volumes of particles with $\dpre\gtrsim24$~\um. We adopt a uniform error of 20\% in the measure of particle volume, mainly driven by smearing, defocus and irregular shape. Also, in some cases the field was crowded, resulting in the wrongful inclusion of some particles that in fact missed the target, passing behind it. This results in the bulk volume being an overestimate of the actual volume hitting the target.

\subsubsection{Post-impact analysis}
\label{sec:methods:analysis:post}

Shortly after every successful experiment, the target was imaged with an optical microscope with a resolution of 0.6~\um~per pixel.  Subsequently, to allow a direct comparison with spacecraft data, the targets were imaged with the COSISCOPE optical microscope of the reference model\footnote{Located at MPS G\"{o}ttingen} of the COSIMA instrument~\citep{Kissel2007}, which has a resolution of 14~\um~per pixel. Two images per target were taken, with grazing-angle illumination by LED lights placed at opposite sides (`M' and `P'; see \citet{Langevin2016} for a description and illustration of this situation). In order to study the morphology of individual deposits, additional imaging was acquired of selected deposits with a Keyence VK-X200K 3D laser scanning confocal microscope ($xy$ pixel size: 0.14~\um), which also measured the height of deposits with a resolution of $\sim0.1$~\um.

Coverage of the target was obtained by taking the per-pixel maximum of the COSISCOPE M and P frames. This image was manually cleaned of artefacts  and the area containing the screws fixing the target to the target holder were masked. The resulting image was thresholded at $1~\sigma$ above the average illumination level of an empty section of the target on the combined M and P frames. The area of individual fragments was calculated; every interconnected particle is considered as one fragment. From this, the size dimension $\dpost$ for fragments was calculated as being the equivalent diameter of a circle with the same area as the fragment. 

The length of the shadow on the separate P and M images provide a measure of the particle height, $h_{\rm post}$. This value is used in calculating the post-impact deposit volume of a single fragment as
\begin{equation}
V_{\rm post} = \frac{\pi}{4}\epsilon{\dpost}^2 h_{\rm post}.
\end{equation}
The parameter $\epsilon$ parametrises the deposit geometry (see \citealt{Hornung2016} and Paper~I). We adopt $\epsilon=0.33$ (a pyramid shape) for all deposits.

We calculated height for selected fragments (around 25\% of the total number) across the deposit size and morphology range, to arrive at an estimate for the typical height-to-base ratio as a function of velocity, monomer size and deposit size. We assume no compaction of the material has taken place during the collisions with the target; see the discussion in Paper~I (Sect. 4.3). As the bulk density of monodispersed material is comparable to or higher than the polydisperse material used in Paper~I, and the bulk density is also similar, we assume compaction can also be neglected in the current study. Combining the total volume of material pre- and post impact, we calculate the mass transfer function $\mathrm{TF}$, being the fraction of the total pre-impact dust mass deposited onto the target in one experiment, as

\begin{equation}
\mathrm{TF} = \left. \sum\limits_{i}^{n_{\rm post}}V_{{\rm post},i} \hspace{3pt} \middle/  \hspace{3pt} \sum\limits_{i}^{n_{\rm pre}}V_{{\rm pre},i} \right.
\label{eq:tf}
\end{equation}

where $V_{\rm pre, i}$ and $V_{\rm post, i}$ are the values of individual particles and fragments, respectively.  

%FIGURE: COLLISION CARTOON
	\begin{figure}
	\begin{center}
	\includegraphics[width=0.9\columnwidth]{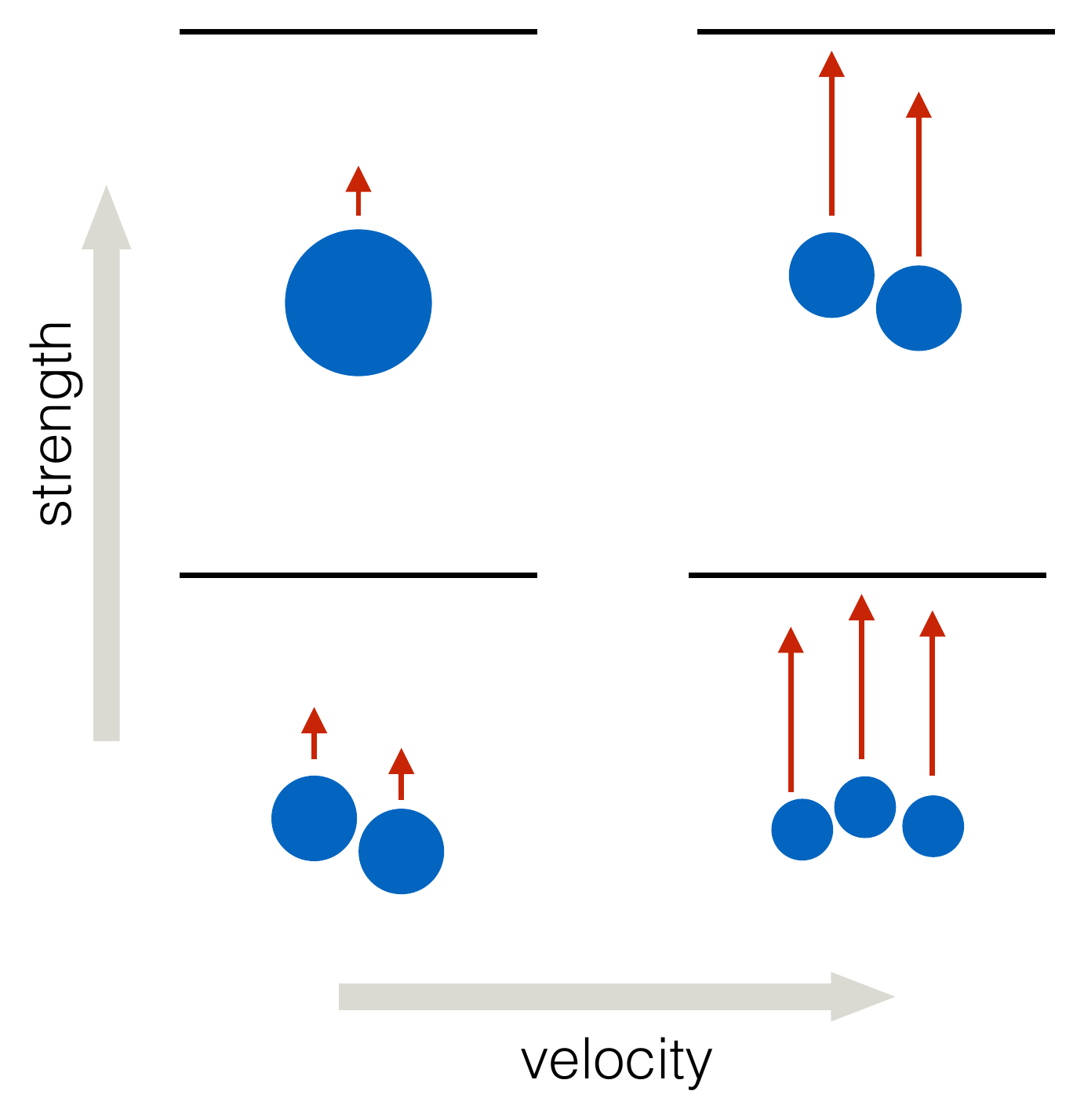}
	\end{center}
    \caption{Qualitative picture of pre-impact particle size distribution in the different experiments.}
    	\label{fig:collisioncartoon}
	\end{figure}

%FIGURE: MATRICES
	\begin{figure*}
	\includegraphics[width=0.7\textwidth]{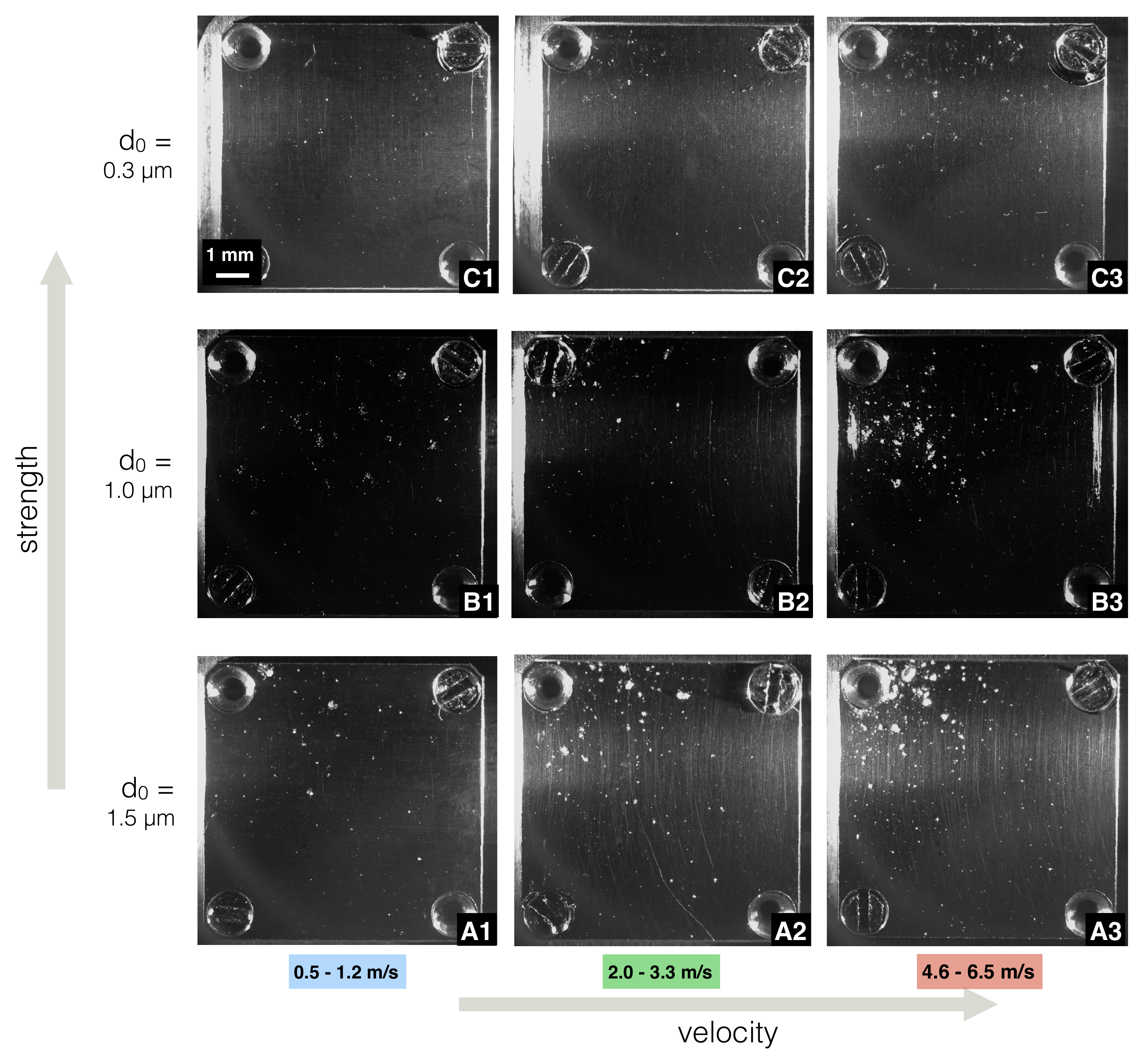}\\
	\includegraphics[width=0.7 \textwidth]{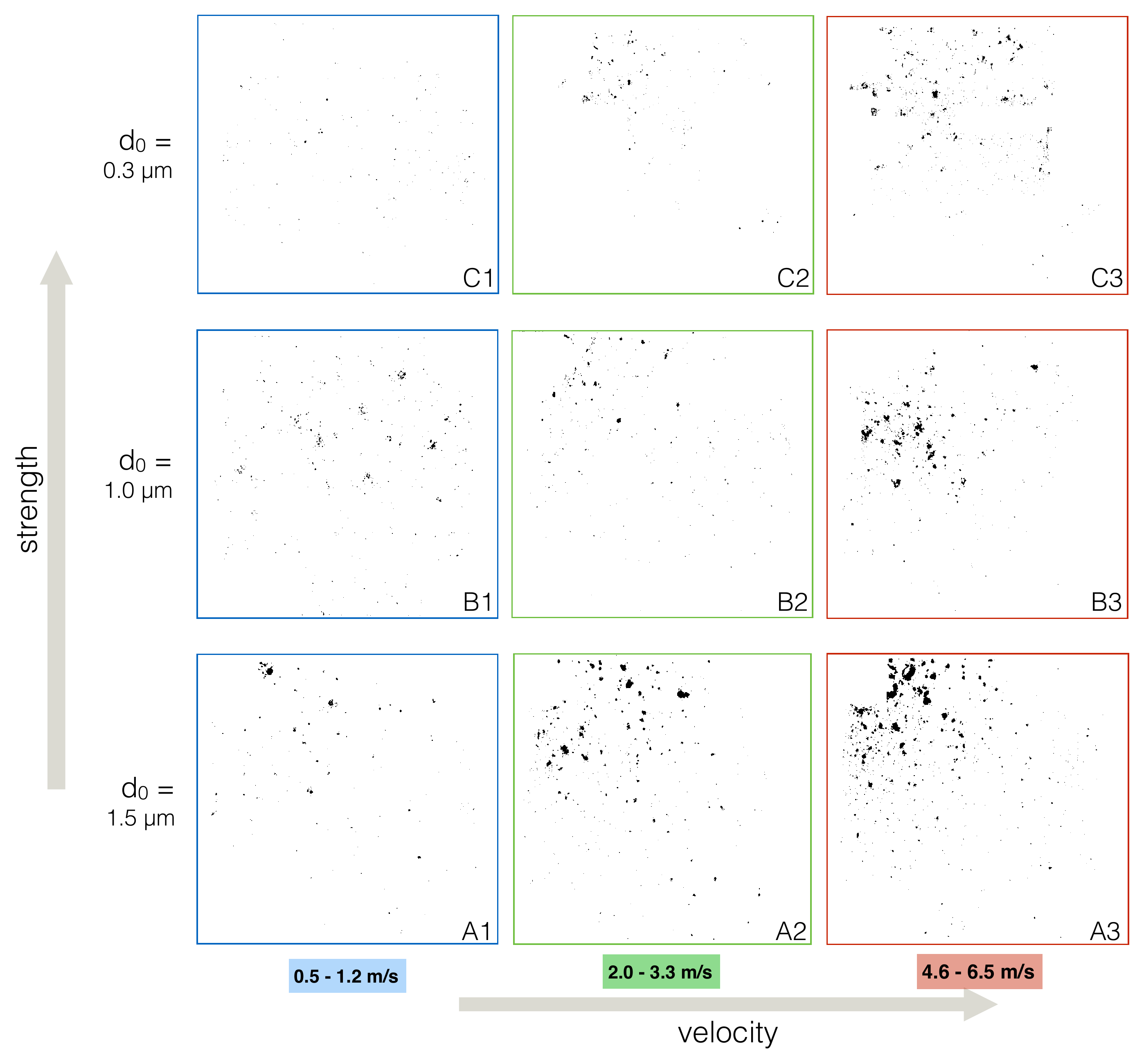}
    \caption{\textit{Top:} Targets after experiments, illuminated from right and left (co-added, see text). \textit{Bottom:} Same as above, with thresholded images highlighting target coverage by dust in black.}
    	\label{fig:matrices}
	\end{figure*}
	
%FIGURE: CUMDIST PLOTS 1
	\begin{figure*}
\includegraphics[height=0.8\textwidth]{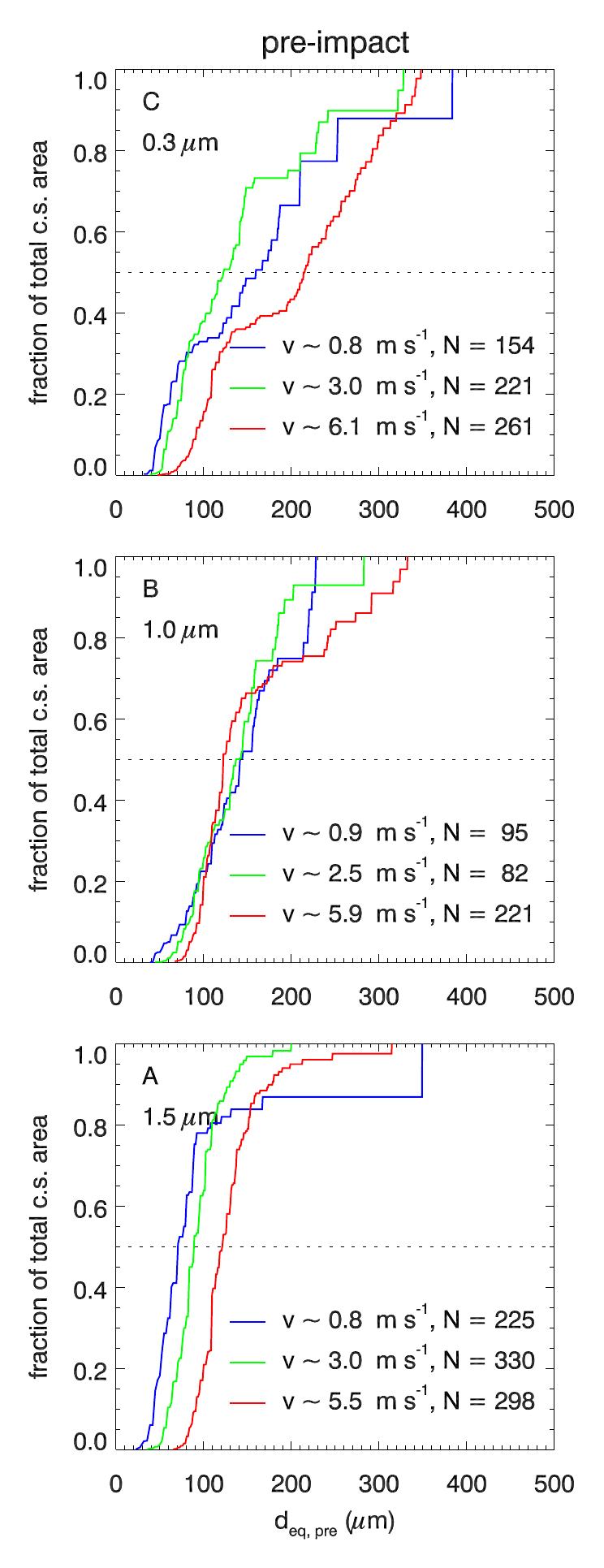} 	
\includegraphics[height=0.8\textwidth]{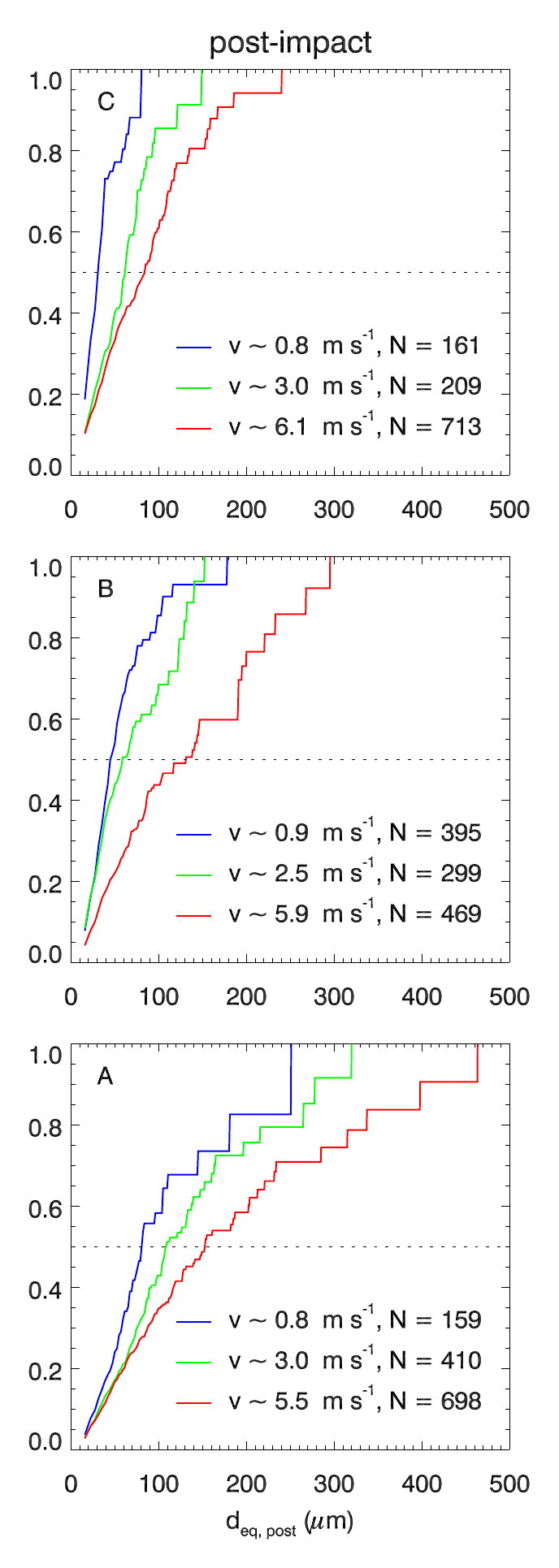}
	\caption{Cumulative size distribution for pre-impact (left column, particles) and post-impact (right column, fragments) sizes, normalised to the total cross-sectional area. Tensile strength decreases per row from top to bottom, as monomer size increases; capital letters A, B and C correspond to experiments listed in Tab.~\ref{tab:experiments}. Note that the absolute value of $V_{\rm post, total}$ (right column), is up to a factor 10--100 lower than $V_{\rm pre, total}$ (left column) (also see middle right panel).}
    	\label{fig:mt_combined}
	\end{figure*}

% TABLE: EXPERIMENTS
\begin{table*}
	\centering
	\caption{Summary of experiments and results. Columns, from left to right: experiment label; monomer diameter; velocity range of particles; number of particles identified on pre-impact movie stills, above threshold (see text); power law index of particle size distribution; number of post-impact fragments on target above threshold (see text); power law index of fragment size distribution; mass transfer function.}
	\label{tab:experiments}
	\begin{tabular}{cccccccc} % four columns, alignment for each
		\hline
		& & \multicolumn{3}{c}{\textit{pre-impact}}  & \multicolumn{3}{c}{\textit{post-impact}} \\
		\\
		     Experiment        &  $d_0$ & Velocity range &  \#particles & $\alpha_{\rm pre}$ &  \#fragments & $\alpha_{\rm post}$ & TF \\
                                &    (\um)  & (m s$^{-1}$)  & ($\dpre~>24$~\um) &  & ($\dpost~>16$~\um) & & (\%) \\
\hline
A1 	& 1.5 & $0.4-1.2$ & 225 & $-1.2 \pm 0.2 $ & 159 & $-1.9 \pm 0.2$  & 1.4 $\pm$ 0.3     \\
A2 	& 1.5 & $2.7-3.3$ & 330 & $-1.3 \pm 0.2 $ & 410 & $-1.6 \pm 0.2$  & 4.8 $\pm$ 0.9     \\
A3 	& 1.5 & $5.7-6.5$ & 298 & $-0.4 \pm 0.1 $ & 698 & $-1.6 \pm 0.2$  & 7.4 $\pm$ 1.5    \\
B1 	& 1.0 & $0.5-1.1$ & 95  & $-0.7 \pm 0.2 $ & 395 & $-2.6 \pm 0.2$  & 0.35 $\pm$ 0.07   \\
B2	& 1.0 & $2.1-2.9$ & 82  & $-0.9 \pm 0.2 $ & 299 & $-1.8 \pm 0.2$  & 0.83 $\pm$ 0.17   \\
B3	& 1.0 & $5.5-6.3$ & 221 & $-2.0 \pm 0.2 $ & 469 & $-1.6 \pm 0.2$  & 2.0 $\pm$ 0.4     \\
C1	& 0.3 & $0.5-1.1$ & 154 & $-2.3 \pm 0.2 $ & 161 & $-2.9 \pm 0.2$  & 0.02 $\pm$ 0.02   \\
C2	& 0.3 & $2.6-3.2$ & 221 & $-1.8 \pm 0.2 $ & 209 & $-2.3 \pm 0.2$  & 0.21 $\pm$ 0.04  \\
C3      & 0.3 & $5.4-5.6$ & 261 & $-3.4 \pm 0.3 $ & 713 & $-2.0 \pm 0.2$  & 0.25 $\pm$ 0.05  \\
\hline
	\end{tabular}
\end{table*}

	%FIGURE: CUMDIST PLOTS 2
	\begin{figure}
\includegraphics[width=\columnwidth]{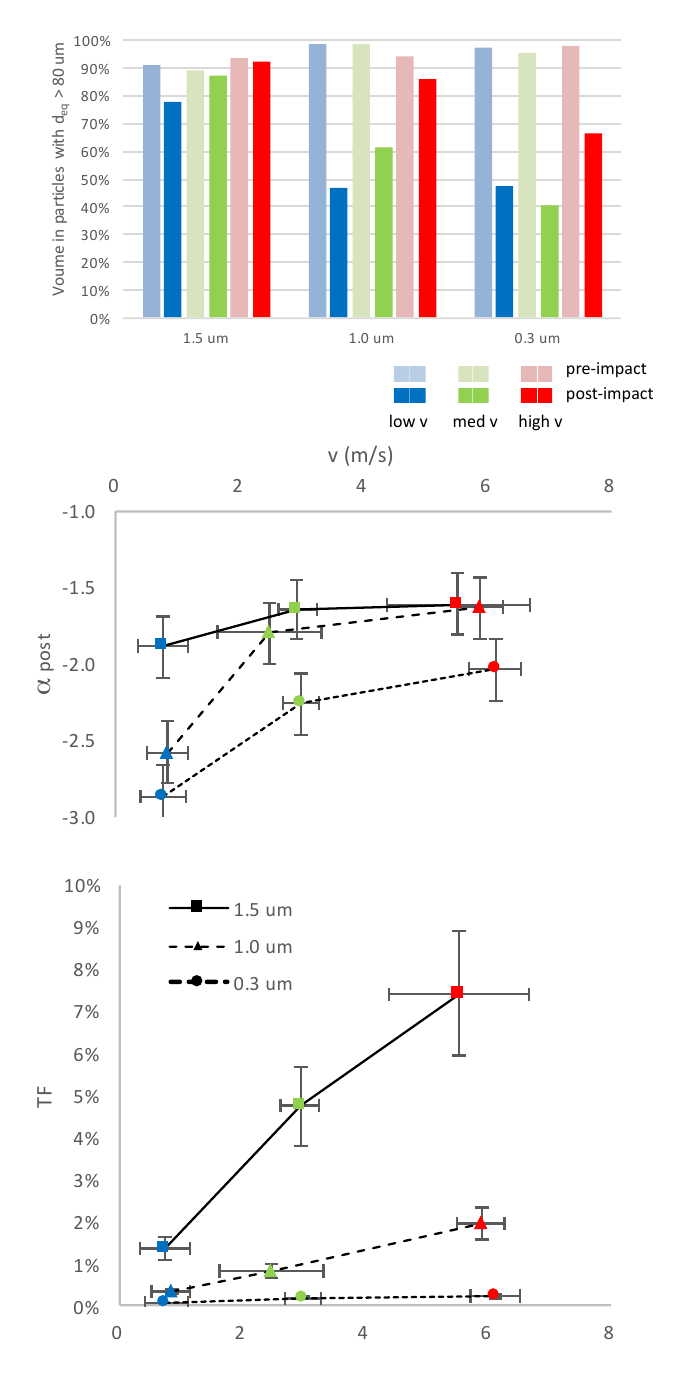}\\ 	
	\caption{Post-impact analysis. \textit{Top:} Fraction of total volume (mass) contained in large (i.e. $\deq>80$~\um) particles (pre-impact) / fragments (post-impact). \textit{Middle:} Power law index $\alpha_{\rm post}$ of the post-impact fragment size distributions related to monomer size and impact velocity. \textit{Bottom:} Mass transfer function $\mathrm{TF}$ related to monomer size and impact velocity. The estimated error for both $\alpha_{\rm post}$ and $\mathrm{TF}$ is 20\%, see Sect.~\ref{sec:methods:analysis:pre}.}
    	\label{fig:mt_combined2}
	\end{figure}

				%FIGURE: MT_TENSILE 
	\begin{figure}
	\begin{center}
	\includegraphics[width=\columnwidth]{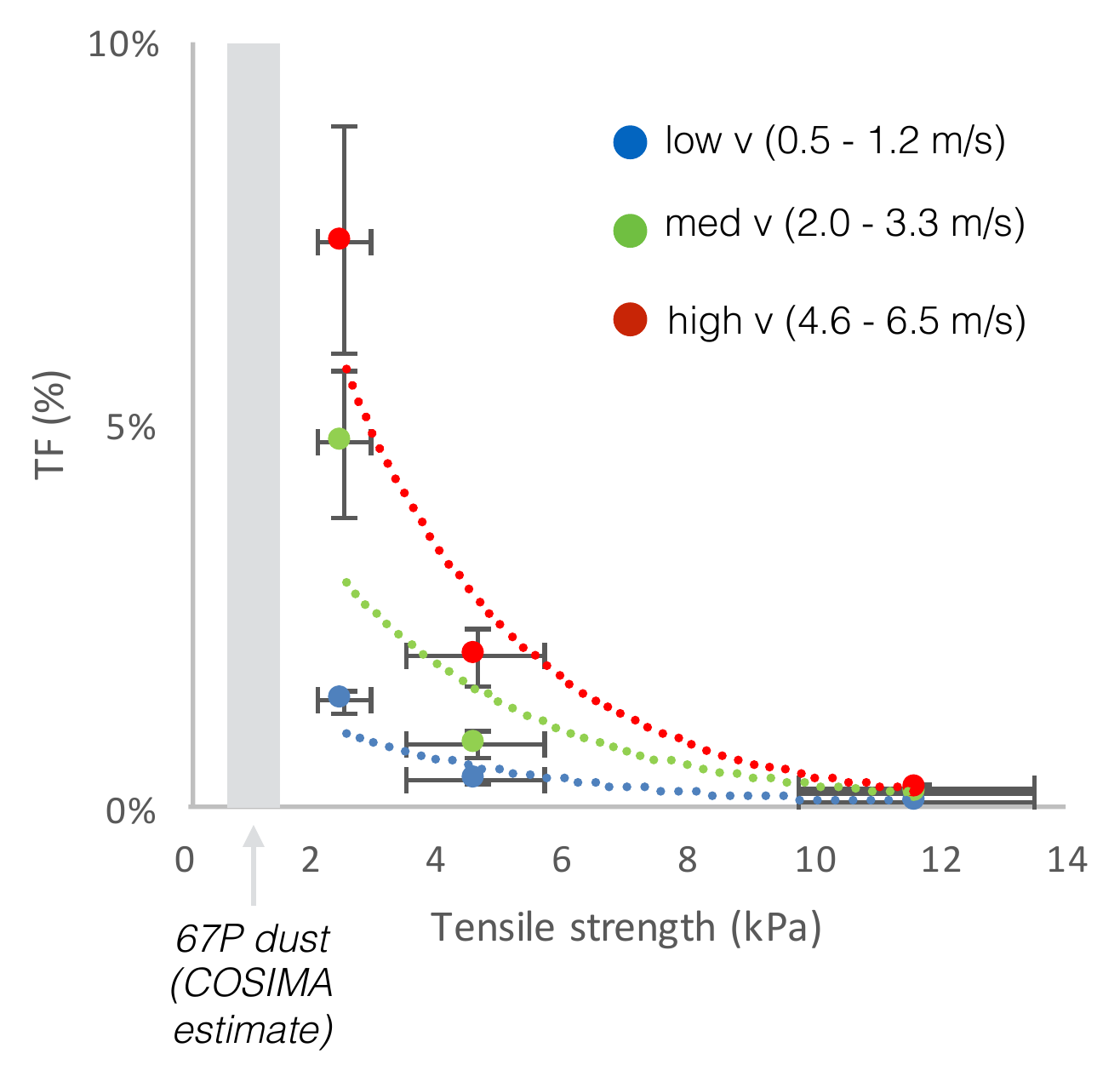}
	\end{center}
    \caption{Mass transfer function measured in experiments related to tensile strength as corresponding to monomer size (see text). The 67P dust strength estimate is from \citet{Hornung2016}. The estimated error for $\mathrm{TF}$ is 20\%, see Sect.~\ref{sec:methods:analysis:pre}.}
    	\label{fig:mt_tensile}
	\end{figure}

%\clearpage
%SECTION: RESULTS
\section{Results}
\label{sec:results}

We present the results of nine experiments, one for every combination of monomer size $d_0=(0.3,1.0,1.5)$~\um~and $v$ (low: $\sim1$~\ms, medium: $\sim2.5$~\ms, high: $\sim6$~\ms). See Tab.~\ref{tab:experiments} for a summary of the analysis. 

Upon analysis of the pre-impact movies, the breakup of particles upon launch caused the material with low tensile strength (large monomer size) to contain relatively more small fragments compared to the high-strength material (small monomer size). This is schematically displayed in Fig.~\ref{fig:collisioncartoon}. Note that in all experiments, over 90\% of the pre-impact mass is contained in particles larger than 80~\um~(see below).

Post-impact images of the nine targets are displayed in Fig.~\ref{fig:matrices} (top panel). The bottom panel of this figure displays thresholded images of these targets, which highlight the coverage with dust fragments (in black). These were used to calculate the area (and hence $\deq$) of individual fragments. The height of $\sim25$\% of the on-target fragments was measured, and averaged over fragments smaller and larger than $80$~\um. These values were used accordingly to calculate the total post-impact volume. The values lie in the range $h/\dpost\sim0.1-0.2$ and decrease with size, tensile strength and velocity, similar to what was measured for the deposits in Paper I.

Upon examining the movies of the impact, we observe the same three classes of collisions (sticking, bouncing, fragmenting) as were identified in Paper~I. At velocities below the fragmentation barrier (see Sect.~\ref{sec:methods:analysis:pre}), small particles ($\lesssim80$~\um), stick, leaving a single deposit, while larger particles bounce off the target, leaving a shallow footprint of small fragments. Towards smaller monomer sizes (higher tensile strength), large particles that bounce leave shallow footprints whose area is more sparsely covered, or no deposit remains at all.  At velocities above the fragmentation barrier, a pyramid-shaped deposit, surrounded by a scatter field of smaller monomers, is left on the target. Towards smaller monomer sizes (higher tensile strength values), individual fragments increasingly bounce off the target, resulting in a smaller value of $\mathrm{TF}$. These trends are further illustrated by Fig.~\ref{fig:laser_collage} in the Appendix, which displays the morphologies of selected deposits for all experiments. From this figure, an empirical `fragmentation barrier' in strength-velocity space may be deduced; a quantitative determination of this line is beyond the scope of this paper.

The pre- and post-impact analyses are summarised in Figs.~\ref{fig:mt_combined}~and~\ref{fig:mt_combined2}. Fig.~\ref{fig:mt_combined} displays pre- and post-impact fragment size distributions (normalised to 1). The top two graphs in Fig.~\ref{fig:mt_combined2} summarise the trends observed in the mass distributions. The histogram on the top panel shows the fraction of the pre- and post-impact mass contained in large ($>80$~\um) particles (pre-impact) and fragments (post-impact) respectively. In all experiments, more than 90\% of the measured pre-impact mass is contained in particles larger than 80~\um~(4 pixels across on the movie image). In the post-impact size distribution, however, a clear trend is visible. Material with lower tensile strength leaves relatively more large fragments on the target. Velocity has a similar effect: at higher velocities, larger deposits are made. 

This effect is also reflected by the graph in the middle panel, which shows the power law index $\alpha_{\rm post}$ of the post-impact size distribution, defined as 
\begin{equation}
N \propto \deq^{\alpha, {\rm post}}, 
\label{eq:powerlaw}
\end{equation}
measured for particles/fragments between $30>\deq>300$~\um, binned with logarithmic intervals of 0.15, similar to \citet[][and references therein]{Merouane2016, Merouane2017}. In general, a steep power law indicates the presence of more small fragments. Fig.~\ref{fig:mt_combined2} (middle panel) shows that in our experiments, low velocity and high strength lead to relatively more smaller fragments on the target. This reiterates the trend that was observed from the histogram in the top panel of this figure. 

The bottom graph shows that the mass transfer function $\mathrm{TF}$ increases with increasing impact velocity and increasing monomer size, hence decreases with increasing material strength. $\mathrm{TF}$ does not exceed 10\% in any of the experiments, and is less than 1\% in some of the low-velocity and high-strength experiments. We observe that higher values of $\alpha_{\rm post}$ are observed in experiments with higher values of $\mathrm{TF}$, which emphasises that the mass is dominated by large deposit fragments. 

The trends observed in $\mathrm{TF}$ and $\alpha_{\rm post}$ imply that when attempting to retrieve pre-impact characteristics from post-impact mass distributions, a degeneracy exists between velocity and tensile strength. We will discuss this further in Sect.~\ref{sec:discussion:degeneracy}.

%SECTION: DISCUSSION
\section{Discussion}
\label{sec:discussion}

\subsection{Factors that influence mass transfer}
\label{sec:discussion:degeneracy}

The main finding of this series of experiments is that, within the parameter space considered, mass transfer efficiency increases with impact velocity and decreases with tensile strength. This is consistent with earlier work by \citet{Guttler2010}; see also the discussion in Paper~I~(Sect.~4.1). At the values considered in this study, even at the softest material ($\sim3$~kPa) at highest velocities ($\sim6$~\ms), $\mathrm{TF}$ does not exceed 10$\%$. Furthermore, our experiments show that multiple parameters influence the appearance of deposits in a similar way. In this subsection we discuss these factors. 

The influence of impact velocity and tensile strength on mass transfer efficiency is similar. This degeneracy is not lifted by considering other observables, like the relative fragment size distribution, as this has a similar scaling relation with both parameters. A closer examination of the individual deposits, however, may prove fruitful in determining its tensile strength or impact velocity, for example by relating this to the height-to-base ratio of individual fragments. Such an in-depth analysis is beyond the scope of this paper.

In the imaging data acquired by the Rosetta mission, just the post-impact situation is known to us. If one were to retrieve the strength and velocity of the particle(s) prior to their impact on the target, one is met with various complications. First of all, the degeneracy described above makes it difficult to isolate the effect of impact velocity. Another complication that inhibits the retrieval of the properties of a parent particle is the unknown pre-impact size distribution. It is expected that some of the particles that entered the instrumentation, were broken up upon collision with the instrument funnel \citep{Merouane2016}. The effects of these funnel collisions may be studied by simulating them in a future series of experiments. 

\subsection{Comparison to cometary dust}
\label{sec:discussion:comparison}

Fig.~\ref{fig:mt_tensile} displays the relation between mass transfer and tensile strength found in our experiments. It can be seen that mass transfer steeply increases towards strengths below 4 kPa. While measurements at lower strengths are not available, it is likely that of the dust that initially hit the COSIMA targets during the Rosetta mission, more than half has not ended up there following a bouncing and/or fragmenting collision. 

Some considerations should be taken into account when extrapolating our experimental results to cometary dust. Firstly, the monodisperse material used in our experiments is not expected to occur in nature. However, it has a clear dependence of strength on monomer size, and allows us to study this parameter. \citet{Langevin2017} show that the albedo of cometary dust is much lower than our sample material. As the contrast with the background of our material is higher than it would have been with darker material, this may have resulted in a systematic overestimate of post-impact fragment sizes. 

Furthermore, we observe that the SiO2 material breaks up into tiny fragments with sizes below the COSISCOPE resolution limit, brightening up neighboring pixels in the images. For cometary dust, in-situ `experiments' of the breakup of cometary dust by impact or electrostatic forces due to charging show that the dust particles do break up into elements in the $10-50$~\um~size range with dark areas in between, without a hint for blurring or an increase in reflection due to partial filling of the instrument pixel with reflections of tiny particles below the resolution limit \citet{Hilchenbach2017, Langevin2016}.

These differences should be considered when directly comparing experimental results with COSIMA data. For a further detailed discussion on the similarity to cometary material, we refer to the Discussion in Paper I (Sect.~4.2). 

Analysis of Rosetta data has suggested the presence of either two different types of dust populations: a more compact component and a fluffy component, or a more uniform composition with one component. In the two different type hypothesis, the compact component constitutes with mass density~$\rho_{\rm b} \sim 10^3$~\kgm~\citep[][]{Rotundi2015, Fulle2016b, Fulle2017}, and a filling factor $\phi>0.1$. A second, low-density component ($\rho_{\rm b}\sim1$~\kgm) is detected by GIADA \citep{Fulle2015, Fulle2016a}. This component is hypothesized to consist of the fractal dust particles detected by MIDAS~\citep{Bentley2016, Mannel2016}, but a limited amount of data is available to confirm this hypothesis. The comet formation model proposed by \citet{Blum2017} also includes the fluffy component to be present in between the more compact particles. \citet{FulleBlum2017} conclude that the presence of this fluffy component suggests that no collisions at velocities higher than 1~\ms~have taken place during formation of the comet.

The COSIMA dust collection observations can be modelled with one dust particle population with a tensile strength due to van der Waals forces of $\sigma\sim~1$~kPa and mass density~$\rho_{\rm b} \sim 1-4\times10^2$~\kgm~for particles in the 60--300~\um~size range \citep{Hornung2016}, a filling factor down to $\phi\sim 0.1$, based on optical observations \citep{Langevin2017} and mechanical parameter analysis~\citep{Hornung2016}. The rationale for the later dust particle model is that particles classified prior as `compact' after impact can be fragmented by Lorentz forces due to in-situ charging in the instrument and therefore do not constitute a separate dust particle class \citep{Merouane2016, Langevin2016, Hilchenbach2017}. 

It is intriguing to find out whether an application of our results to COSIMA can shed light on the strength of cometary dust, or the existence of the different dust species.

			%FIGURE: COSIMA DATA
	\begin{figure*}
	\begin{center}
	\includegraphics[width=0.93\columnwidth]{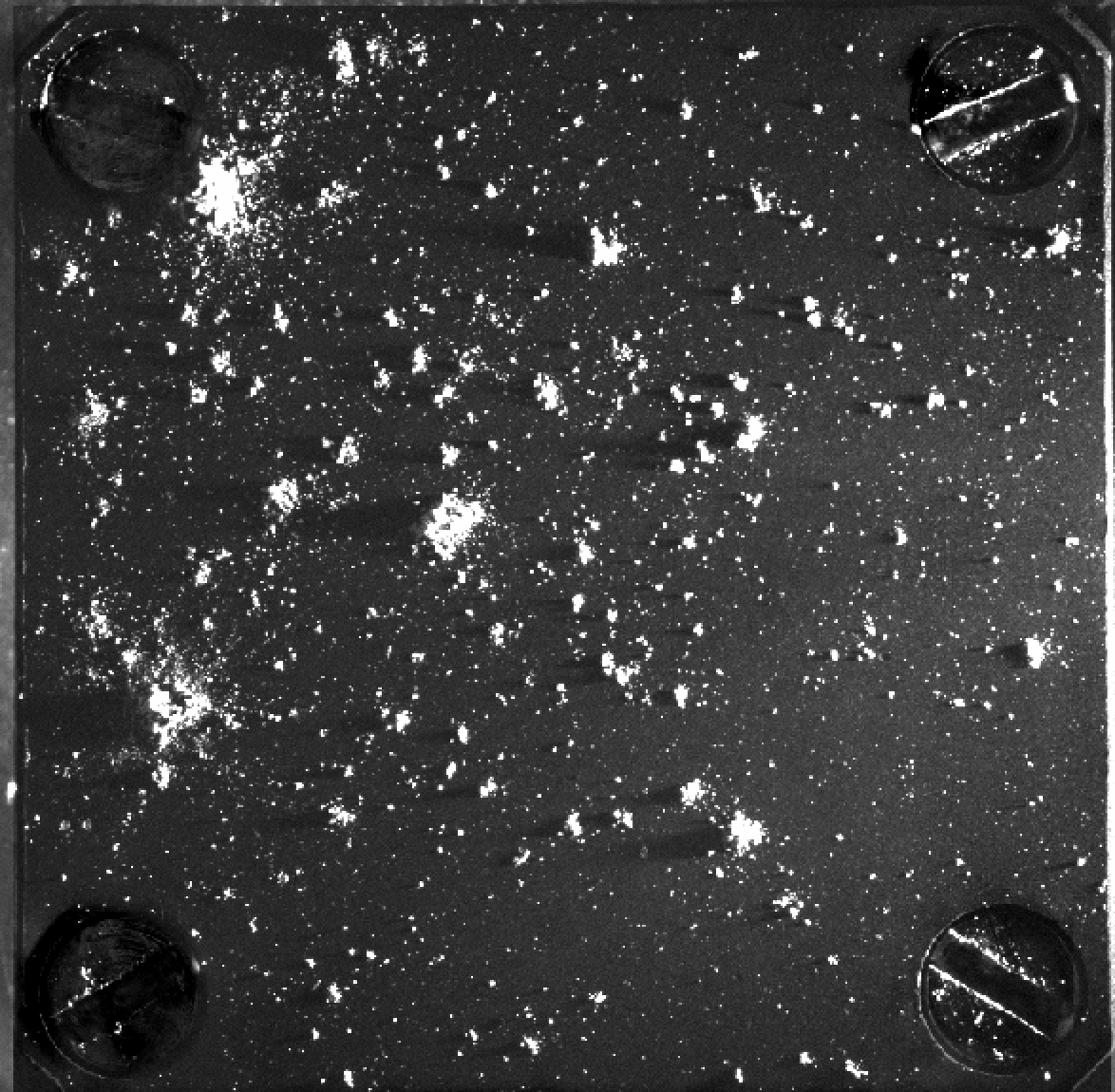}
	\includegraphics[width=0.9\columnwidth]{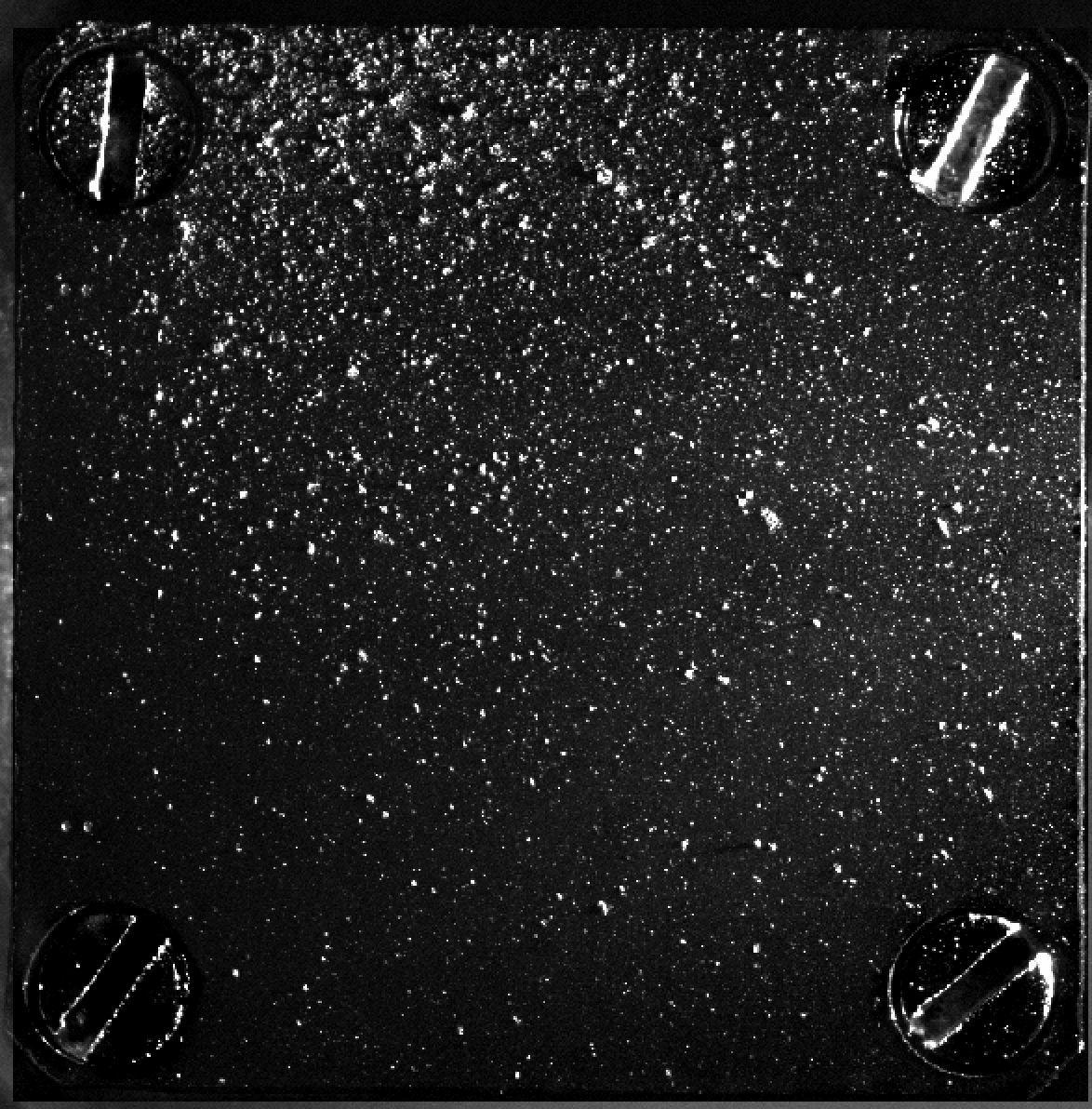}
	\end{center}
    \caption{Two different COSIMA targets; 2D1 (left) was exposed for 7 weeks before perihelion, 1CD (right) was exposed for 19 weeks including perihelion.}
    	\label{fig:cosimadata}
	\end{figure*}

% TABLE: COSIMA TARGETS
\begin{table*}
	\centering
	\caption{Summary of properties of COSIMA targets displayed in Fig.~\ref{fig:cosimadata}.}
	\label{tab:cosima}
	\begin{tabular}{cccccccccc} % four columns, alignment for each
		\hline
		Target & exposure & spacecraft  & \multicolumn{2}{c}{heliocentric} & Collision event & \% of on-target  & spacecraft & heliocentric & \\
		             & dates   & altitude (km) & \multicolumn{2}{c}{distance} &          date       & volume collected & altitude & distance  \\
                                &              &                        & start & end & & during event & (km) & (au) \\
\hline
2D1 & 10/04/2015 -- & 91 -- 321 &  1.89 & 1.55  & 11-12/05 & 73\% & 146 -- 162 & 1.66 (in)\\
     & 27/05/2015 & & (inbound) & (inbound) & 16-17/05 & 24\% & 126 -- 133 & 1.62 (in)\\
      \hline
1CD & 30/05/2015 -- & 153 -- 1502 & 1.53 & 1.41 & 31/07-01/08 & 88\% & 203 -- 215 & 1.25 (in)\\ %1.24 in file header
        & 07/10/2015 & & (inbound) & (outbound) \\
 \hline
%Total & 110 & $30-410$ & $0.3-6.0$ \\
%\aline
	\end{tabular}
\end{table*}
			%FIGURE: COSIMA DATA ANALYSIS
	\begin{figure}
	\begin{center}
	\includegraphics[width=0.82\columnwidth]{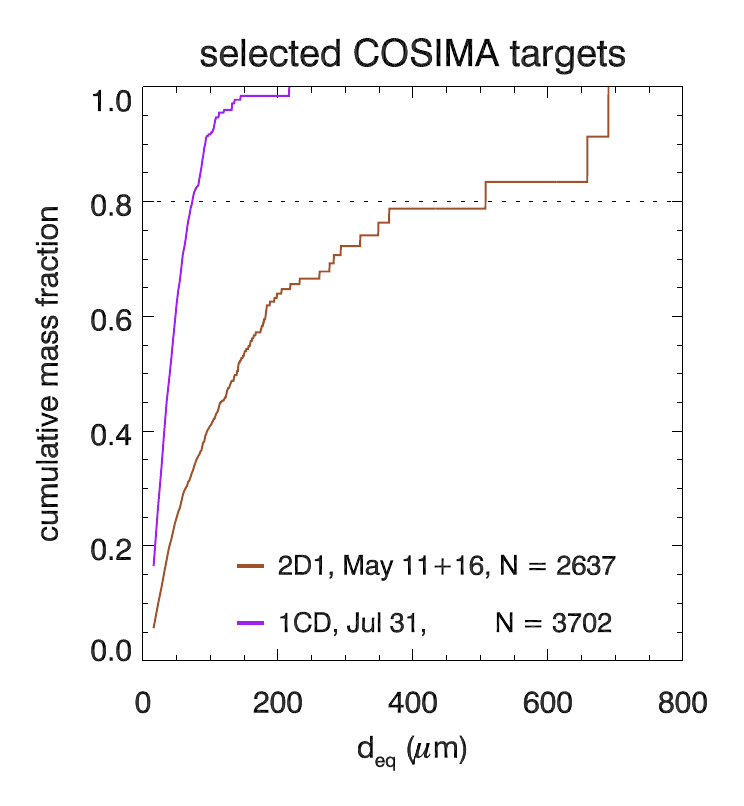}
	\end{center}
    \caption{Fragment size distribution for selected COSIMA targets. Similar to Fig.~\ref{fig:mt_combined} (right column) but applied to the targets displayed in Fig.~\ref{fig:cosimadata}. $N$ corresponds to the total number of fragments on target.}
    	\label{fig:cosima_analysis}
	\end{figure}

%SECTION: APPLICATION TO COSIMA DATA
\subsection{Application to Rosetta/COSIMA measurements}
\label{sec:discussion:cosima}

As an illustration of the application of these experiments, we applied our method to Rosetta data. We compare two COSIMA targets that have captured multiple particles. Fig.~\ref{fig:cosimadata} shows stray-light corrected images (P illumination) of the two targets. Some properties of the targets and their collection periods are listed in Tab.~\ref{tab:cosima}. The targets have been exposed for 7 weeks (2D1) and 19 weeks (1CD). During the second collection period, at least two major outbursts (May 6 and August 15, 2015) and several smaller events have taken place \citep{Feldman2016, Knollenberg2016, Vincent2016, Pajola2017}. However, the bulk of the fragments on these targets were collected during  single collision events outside the time window of the major outburst events, as was identified by \citet[][Tab.~A1]{Merouane2017}. From this analysis, it can be inferred that these collision events took place within $\sim24$~hour time windows 3 months (2D1: May 11-12 and 16-17, 2015) and 2 weeks (1CD, 31 Jul - 1 Aug, 2015) before perihelion. The spacecraft speed relative to the comet was less than 0.1~\ms at all times during these collection events. The major differences in the circumstances were that during the second collection period the spacecraft was both 1.5 times further away from the comet surface, and 1.3 times closer to the sun. 

\citet{Merouane2017} find that the power law index of the fragment size distribution increases during outburst events, due to either higher velocities or material with lower tensile strength. While there is no evidence that the collision events have coincided with outburst events, a possible effect contributing to the abundance of small fragments on target 1CD is that in the period around an outburst event, higher pressure may lead to breakup into smaller aggregates. These can be lifted more easily because of the inverse tensile strength relation of aggregate packings \citep[see][]{SkorovBlum2012}. When we apply the same method as described in Sect.~\ref{sec:results}, the power law index for 2D1 is $\alpha \sim -1.9$, and for 1CD $\alpha \sim -3.6$; see Fig.~\ref{fig:cosima_analysis}. In other words, target 1CD contains a relative high amount of small fragments. If the tensile strength of both dust particles is equal, this implies a higher impact velocity for period 1CD. Conversely, if velocities were similar, the 1CD particle may have been of a stronger dust species. 

A more likely scenario is that in both cases, a particle collided with the funnel wall and broke up into smaller parts. For 1CD, however, the particle may have hit at higher velocity, causing breakup into more and smaller fragments (consistent with our pre-impact analysis). These fragments subsequently impacted at lower velocities. Our result thus shows that the dust on the two COSIMA targets presented are consistent with originating from particles with a single stength. The significant difference in the distribution of particles on their surface may thus be explained by different impact velocities. 

%SECTION: SUMMARY AND CONCLUSIONS
\section{Summary and Conclusions}
\label{sec:conclusions}

Aiming to simulate the circumstances of dust collection in the COSIMA and MIDAS instruments on the Rosetta spacecraft, we performed nine experiments, in which we collided SiO$_2$ aggregates in the size range 30~--~400~\um~with a target surface at impact velocities of 0.5~--~6.5~\ms. Three different monodisperse monomer sizes were used, resulting in tensile strengths in the range 3~--~12~kPa. We used COSIMA targets, and also COSISCOPE imaging to obtain a like-for-like comparison with COSIMA data. Our main conclusions are:
\begin{itemize}
\item[(i)] The transfer function increases with increasing velocity and increasing monomer size (hence, decreases with increasing tensile strength).
\item[(ii)] No more than 10\% of the impacting material is transferred during collisions on the target surface. This fraction drops to below 1\% for low velocity and high tensile strength.
\item[(iii)] Material with lower tensile strength leaves relatively more deposits exceeding 80~\um~in size.
\item[(iv)] Extrapolating these results to the strength parameters of cometary dust imply that more than half of the dust that enters COSIMA and MIDAS was lost in the instrumentation.
\end{itemize}

While differences in material properties, unknowns pre-impact and degeneracies between parameters inhibit a direct comparison to COSIMA data, the qualitative results of the experiments may be used to correlate trends found in the fragment distribution patterns. 
In future space missions to comets (e.g. CAESAR sample return mission to 67P, \citealt{Squyres2018}), we suggest to use a different technique (e.g. combination of laser curtain, enhanced imaging capabilities, multiple resolution modes for scanning) that avoids the uncertainty introduced by the impact in the instruments.

\section*{Acknowledgements}

The authors thank Klaus Hornung for useful discussion. This work has been financially supported by NWO, project no. ALW-GO/15-01. The SEM image (Fig.~\ref{fig:samples}) was taken at the Electron Microscopy Center Amsterdam, Academic Medical Center, Amsterdam, The Netherlands.  A.L. is funded by DFG through grant Bl298/24-1. J.B. and B.G. thank the DFG for continuous funding; B.G. thanks the DFG for continous support for the CoPhyLab project (DFG GU 1620/3-1 and BL 298/26-1 / SNF 200021E 177964 / FWF I 3730-N36).

\bibliographystyle{mnras}
\input{footprint2_revised2_clean.bbl}

\clearpage
\appendix
\section{Images of individual dust deposits}

	%FIGURE: LASER COLLAGE
	\begin{figure*}
	\includegraphics[width=0.95\textwidth]{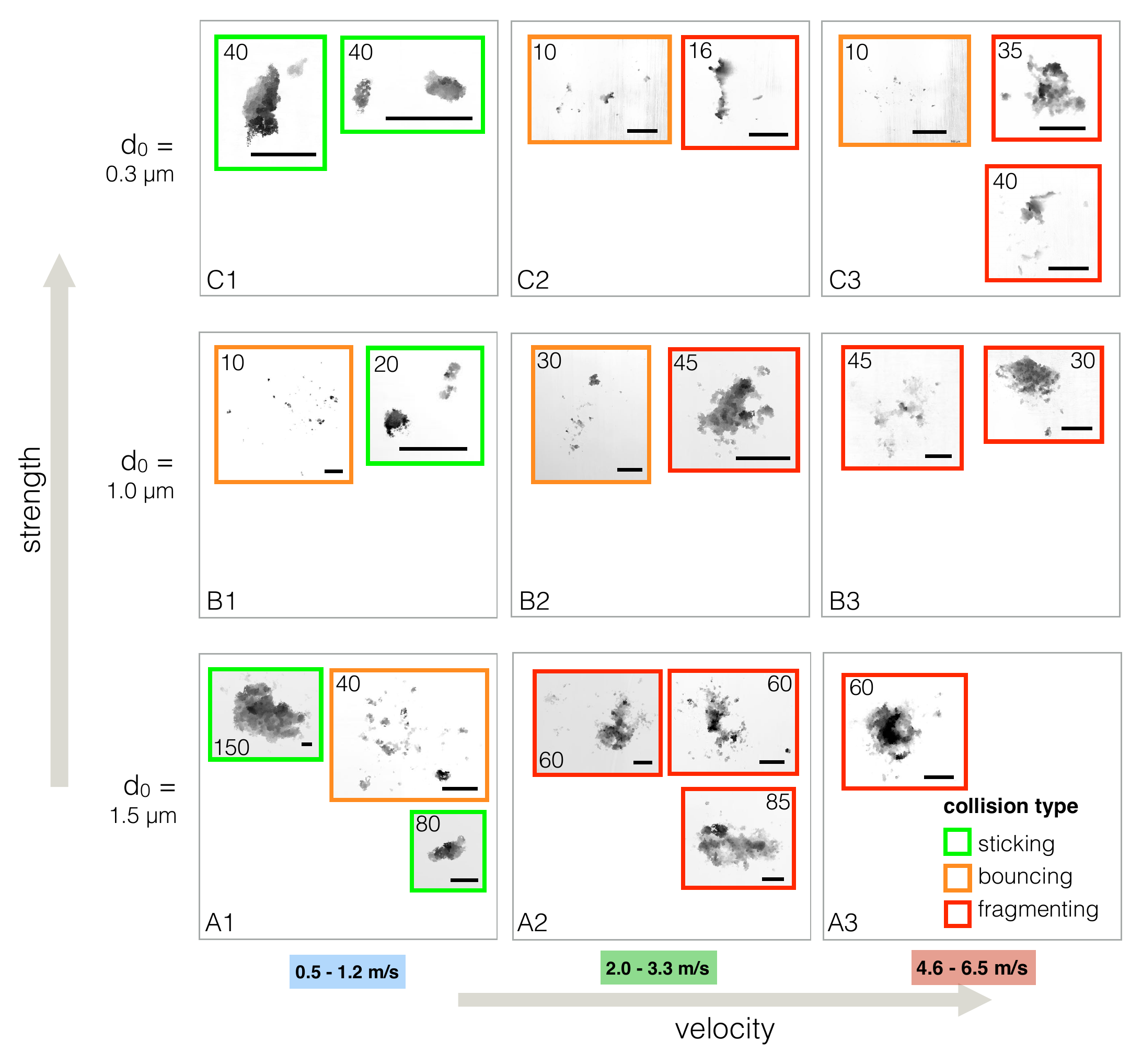}\\
    \caption{Examples of individual deposit morphologies. The images displayed are height maps obtained with a 3D laser scanning confocal microscope. The height scale is normalized to the maximum height of a deposit; numbers within boxes correspond to the maximum height level (in~\um) of the deposit. Frame colours indicate the type of collision that led to the deposit. All scale bars indicate 100~\um.}
    	\label{fig:laser_collage}
	\end{figure*}

%%%%%%%%%%%%%%%%% APPENDICES %%%%%%%%%%%%%%%%%%%%%

%\section{Some extra material}

%If you want to present additional material which would interrupt the flow of the main paper,
%it can be placed in an Appendix which appears after the list of references.

%%%%%%%%%%%%%%%%%%%%%%%%%%%%%%%%%%%%%%%%%%%%%%%%%%

% Don't change these lines
\bsp	% typesetting comment
\label{lastpage}
\end{document}